\documentclass[onecolumn,sort&compress,numbers]{els-mrw} 

\usepackage{amsmath,amssymb,amsfonts,amsthm,makeidx,graphicx}
\usepackage{txfonts}
\usepackage{helvet}
\usepackage{hyperref}
\usepackage{cancel}
\usepackage{xcolor}

\renewcommand{\vec}[1]{\ensuremath{\mathbf{#1}}}

\begin{document}

\chapter{Radiative corrections to weak interaction processes}\label{chap1}

\author[1]{Chien-Yeah Seng}%

\address[1]{Department of Physics and Astronomy, University of Tennessee, Knoxville, Tennessee 37996, USA}


\maketitle

\begin{abstract}[Abstract]
The search for physics beyond the Standard Model in low-energy weak interaction processes requires a precise knowledge of the Standard Model background at tree- and loop-level. In many such cases, radiative corrections represent one of the major sources of theory uncertainty due to the non-perturbative structures of nucleon or nuclei. In this review, I discuss several modern strategies based on dispersion relation, lattice gauge theory and nuclear many-body calculations to pin down the hadronic and nuclear uncertainties in the radiative corrections to low-energy weak interaction processes such as beta decays and parity-violating electron-nucleus scattering.   
\end{abstract}

\begin{keywords}
 	Standard Model\sep radiative corrections \sep Generalized Compton tensor \sep box diagrams \sep beta decay \sep parity-violating electron scattering \sep dispersion relation \sep Lattice QCD \sep \textit{ab initio} methods \end{keywords}

\section*{Key points}
\begin{itemize}
	\item Non-perturbative strong interaction physics in radiative corrections resides in the generalized Compton tensor $T^{\mu\nu}$.
	\item Dispersion relation connects $T^{\mu\nu}$ to structure functions that can be inferred from experiments, leading to the data-driven analysis of radiative corrections.
	\item For meson and single nucleon, $T^{\mu\nu}$ can also be computed with lattice QCD.
	\item In nuclear systems, the impact from the low-energy part of $T^{\mu\nu}$ can be computed either in terms of nuclear Green's function or nuclear matrix elements of effective two-body potentials, both with cutting-edge nuclear \textit{ab initio} techniques.
\end{itemize}

\section{Introduction}
Understanding the fundamental interactions between the elementary constituents in the universe is one of the ultimate goals in basics sciences. At the present, the Standard Model (SM) of particle physics~\cite{Glashow:1961tr,Weinberg:1967tq,Salam:1968rm} provides the best description of this subject. In this theory, the spectrum of elementary particles consists of six leptons ($e,\mu,\tau,\nu_e,\nu_\mu,\nu_\tau$), six quarks ($u,d,s,c,b,t$) and their anti-particles, which interact through the exchange of gauge bosons: gluons from the $SU(3)_c$ gauge group that transmit the strong force, $\gamma$, $W$ and $Z$ in the $SU(2)_L\times U(1)_Y$ gauge group that transmit the electromagnetic (EM) and weak (electroweak) forces. The SM is arguably one of the most successful theory in physics ever, which is known to pass almost all experimental tests conducted on the earth. However, despite such huge success, it does not account for many important observed phenomena at the largest scale, such as the existence of dark energy~\cite{Aghanim:2018eyx,Riess:1998cb,Perlmutter:1998np}, dark matter~\cite{Aghanim:2018eyx,Simon:2019nxf,Salucci:2018hqu,Allen:2011zs}, and the matter-antimatter asymmetry of the universe~\cite{Aghanim:2018eyx,Mossa:2020gjc}. Therefore, it is now a consensus that the SM is an incomplete theory that needs to be extended.
 
 The search for physics beyond the Standard Model (BSM), or new physics in short, can proceed along a few different directions. One major direction is the ``energy frontier'': by colliding highly energetic particle beams in colliders, one hopes to create new particles not included in the list of particles above. The Large Hadron Collider (LHC), for instance, serves this purpose at TeV scale. However, after the discovery of Higgs boson in 2012 that completed the SM particle spectrum~\cite{Chatrchyan:2012ufa,Aad:2012tfa}, there is so far no definitive evidence of any new particle being observed in collider experiments. Therefore, further progress along this direction will inevitably involve constructions of even more energetic particle colliders. Another direction, known as ``precision frontier'' (or ``intensity frontier'', which I will use interchangeably), proceed as follows: experiments are done at a much lower energy scale to measure certain observable to extremely high precision. With this, one compares the result of the experimental measurement and the theory prediction from the SM; any confirmed discrepancy between the two beyond uncertainties would indicate the existence of new physics. Efforts at both frontiers are equally important to provide a more complete picture of new physics.   
 
 An important class of efforts at the precision frontier is the test of ``fundamental symmetries'' in SM, such as charge conjugation (C), parity (P), time reversal (T), and so on. In the SM, these symmetries are only broken by the weak interaction, which is much weaker than the strong and EM interactions. This means, testing how much these fundamental symmetries are broken provides a powerful avenue to probe signals of new physics with reduced SM background. The effective field theory (EFT) description of BSM physics provides a useful guidance for the precision goal of low energy experiments~\cite{Weinberg:1979sa,Wilczek:1979hc,Buchmuller:1985jz,Grzadkowski:2010es,Jenkins:2013zja}: if new physics appears as dimension-six operators, its impact on electroweak observables may scale as $(v_H/\Lambda)^2$, where $v_H\sim 246\text{ GeV}$ is the Higgs vacuum expectation value, and $\Lambda$ is the characteristic energy scale of new physics. This power counting implies that, to probe new physics at $\sim 10\text{ TeV}$ scale in electroweak observables, one generally requires a precision of $10^{-3}-10^{-4}$. This represents a real challenge not only from the experimental side, but also the side in terms of the SM theory predictions of the observable of interest.  
 
 In this article we discuss an important class of SM effects known as EM radiative corrections (RC), that affects various weak interaction-induced processes such as decays and scatterings. These are corrections on top of the tree-level amplitude due to spontaneous emission and re-absorption of virtual photons, as well as the emission of real photons, by the interacting system. At one loop, its affect is of the order $\alpha/\pi\sim 10^{-3}$ or above, which means it needs to be computed to high accuracy to match our precision goal. If photons would just couple with leptons, then the evaluation of RC would be just a multi-loop problem in Quantum Electrodynamics (QED) which can be computed using standard diagrammatic technique. The real issue, however, is that photons also interact with quarks, and the latter form hadrons due to the strong interaction governed by Quantum Chromodynamics (QCD). Therefore, RC represents a mixture between the perturbative QED and the non-perturbative QCD, and the latter is the main source of theory uncertainty. To pin down the hadron and nuclear physics in the RC require fully non-perturbative methods such as dispersion relation, lattice QCD or \textit{ab initio} nuclear many body methods. 
 
 The proper control of hadronic or nuclear uncertainties in RC is known to play a decisive role in the precision tests of the SM. A well-known example is the muon $g-2$ problem, which status has recently changed from having a $4\sigma$ tension with the SM~\cite{Muong-2:2021ojo,Aoyama:2020ynm} to almost consistent with the SM~\cite{Muong-2:2025xyk,Aliberti:2025beg}. This drastic change is due to a new lattice QCD calculation of the leading-order hadronic vacuum polarization, which is an important RC to the EM interaction of muon~\cite{Borsanyi:2020mff}. The same situation may also occur in weak interaction processes. In this article we will focus on two types of processes: (1) charged weak (CW) decays, and (2) parity-violating electron scattering. We will discuss their roles in the test of SM and the review the existing effort in pinning down the RC.
 
 \section{Basic notations and targets of study}
 
 We start by introducing the notations adopted in this article. Since our ultimate goal is to test the predictions of the SM in the electroweak (EM + weak) sector, it is instructive to first write down the SM electroweak interaction Lagrangian between quarks and gauge bosons\footnote{While $J_\text{em}^\mu$ is unambiguous, notice that in different literature $J_W^\mu$ and $J_Z^\mu$ can normalize differently.}:
\begin{equation}
	\mathcal{L}=-\frac{g}{2\sqrt{2}}\left(J_W^\mu W_\mu^++\text{h.c.}\right)-eJ_\text{em}^\mu A_\mu-\frac{g}{2c_w}J_Z^\mu Z_\mu~,\label{eq:Lew}
\end{equation}
where $e,g$ are the EM and weak coupling constants, $c_w=\cos\theta_W, s_w=\sin\theta_W$ with $\theta_W$ the weak mixing angle. The electroweak currents read:
\begin{eqnarray}
	J_\text{em}^\mu&=&e_u\bar{u}_i\gamma^\mu u_i+e_d\bar{d}_i\gamma^\mu d_i\nonumber\\
	J_W^\mu&=&\bar{u}_i V_{ij}\gamma^\mu(1-\gamma_5)d_j\nonumber\\
	J_Z^\mu&=&\bar{u}_i\gamma^\mu(g_V^u+g_A^u\gamma_5)u_i+\bar{d}_i\gamma^\mu(g_V^d+g_A^d\gamma_5)d_i~.
\end{eqnarray}
Here we use the subscript $i$ to denote the generation of quark: $u_{1,2,3}=\{u,c,t\}$ and $d_{1,2,3}=\{d,s,b\}$. $V_{ij}$ is the Cabibbo-Kobayashi-Maskawa (CKM) matrix that mixes different generation of quarks in the CW interaction~\cite{Cabibbo:1963yz,Kobayashi:1973fv}. Meanwhile, $e_u=2/3$, $e_d=-1/3$ are the electric charge of the quarks, and $g_V^q=I_{3L}^q-2e_q s_w^2$, $g_A^q=I_{3L}^q$ are the vector and axial neutral charges of the quarks, with $I_{3L}^q$ the third component of the weak isospin ($I_{3L}^u=1/2$, $I_{3L}^d=-1/2$).   
Similar expressions can be written down in the lepton sector, except that for practical purpose one can take $m_\nu\rightarrow 0$ which renders the flavor mixing matrix irrelevant. 

\begin{figure}[tb]
	\includegraphics[scale=0.3]{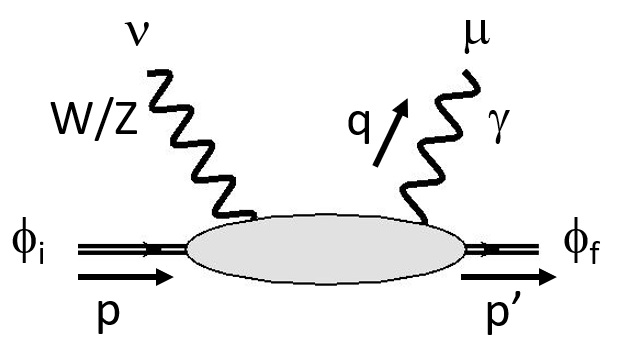}\hfill
	\caption{\label{fig:Tmunu}Diagrammatic representation of the generalized Compton tensor $T^{\mu\nu}$.}
\end{figure}

The amplitude of weak interactions in hadron or nuclear systems at tree level simply consists of the hadronic or nuclear matrix elements of $J_W^\mu$ or $J_Z^\mu$. The RC to tree-level amplitudes involves further insertions of the EM current $J_\text{em}^\mu$. A particularly important matrix structure is the following ``Generalized Compton tensor'':
\begin{equation}
	T^{\mu\nu}(p',p,q)=\int d^4x e^{iq\cdot x}\langle \phi_f(p')|T\{J_\text{em}^\mu(x)J_a^\nu(0)\}|\phi_i(p)\rangle~,\label{eq:Compton}
\end{equation} 
where $J_a^\nu$ can either be the CW current $J_W^\nu$ or the neutral weak (NW) current $J_Z^\nu$, and $\phi_{i,f}$ are the initial and final hadron/nuclear state. This tensor represents the process where the system exchanges two bosons: a photon (couples to $J_\text{em}^\mu$) with an outgoing momentum $q$, and a weak boson ($W$ or $Z$, depends on the current $J_a^\nu$) with an incoming momentum $p'-p+q$ (see Fig.\ref{fig:Tmunu}). In practical applications one often needs to work on;y in the $p'\rightarrow p$ limit (forward limit), where the tensor adopts the following gauge-invariant decomposition:
\begin{eqnarray}
	T^{\mu\nu}(p,q)&=&\left(-g^{\mu\nu}+\frac{q^\mu q^\nu}{q^2}\right)T_1(\nu,Q^2)+\left(p^\mu-\frac{p\cdot q q^\mu}{q^2}\right)\left(p^\nu-\frac{p\cdot q q^\nu}{q^2}\right)\frac{T_2(\nu,Q^2)}{p\cdot q}-i\varepsilon^{\mu\nu\alpha\beta}\frac{q_\alpha p_\beta}{2p\cdot q}T_3(\nu,Q^2)\nonumber\\
	&&+i\varepsilon^{\mu\nu\alpha\beta}\frac{q_\alpha}{p\cdot q}\left[s_\beta S_1(\nu,Q^2)+\left(s_\beta-\frac{s\cdot q}{p\cdot q}p_\beta\right)S_2(\nu,Q^2)\right]+\dots~,\label{eq:Tmunudecompose}
\end{eqnarray}
with $Q^2\equiv-q^2$, $\nu\equiv p\cdot q/M$, $M$ the mass of the external state, and $s^\mu$ is the spin vector of the external state. 

The invariant functions $T_i$, $S_j$ encode all the non-perturbative physics that enter the RC, so their precise determination is the key to reduce the theory uncertainties. Two possible ways to achieve this goal are as follows:
\begin{enumerate}
	\item \textbf{Data-driven approach:} In this approach one first expresses the discontinuity of $T^{\mu\nu}$ with respect to the variable $\nu$ (the energy of the virtual gauge boson) in the physical region ($\nu>0$) as:
	\begin{equation}
		\text{Disc}T^{\mu\nu}\equiv T^{\mu\nu}(\nu+i\epsilon)-T^{\mu\nu}(\nu-i\epsilon)=4\pi W^{\mu\nu}~
	\end{equation}
	The hadronic tensor $W^{\mu\nu}$ is defined as:
	\begin{eqnarray}
		W^{\mu\nu}(p,q)&=&\frac{1}{4\pi}\sum_{X}(2\pi)^4\delta^4(p+q-p_X)\langle f|J_\text{em}^\mu|X\rangle\langle X|J_a^\nu|i\rangle\nonumber\\
		&=&\left(-g^{\mu\nu}+\frac{q^\mu q^\nu}{q^2}\right)F_1(\nu,Q^2)+\left(p^\mu-\frac{p\cdot q q^\mu}{q^2}\right)\left(p^\nu-\frac{p\cdot q q^\nu}{q^2}\right)\frac{F_2(\nu,Q^2)}{p\cdot q}-i\varepsilon^{\mu\nu\alpha\beta}\frac{q_\alpha p_\beta}{2p\cdot q}F_3(\nu,Q^2)\nonumber\\
		&&+i\varepsilon^{\mu\nu\alpha\beta}\frac{q_\alpha}{p\cdot q}\left[s_\beta g_1(\nu,Q^2)+\left(s_\beta-\frac{s\cdot q}{p\cdot q}p_\beta\right)g_2(\nu,Q^2)\right]+\dots~,\label{eq:Wmunu}
	\end{eqnarray}
	where $X$ denotes all possible on-shell intermediate states. 
	The invariant functions $F_i$, $g_j$ are known as ``structure functions''; they depend on on-shell current matrix elements and can in principle be inferred from experimental data, in particular inclusive scattering cross sections. One then constructs a dispersion relation that relates the full invariant amplitudes to the corresponding structure functions. In this way, one converts the theory problem of computing the invariant amplitudes into an experimental problem of measuring the structure functions in the relevant kinematic regions.
	\item \textbf{First-principles theory calculations:} In parallel to the approach above, one may perform pure theory calculations of the invariant amplitudes in $T^{\mu\nu}$, which requires fully non-perturbative treatments of QCD from first principles. By ``first principles'', we mean computational methods that do not involved uncontrolled and non-improvable approximations. When $\phi_{i,f}$ are mesons or single nucleon, lattice QCD represents the only applicable method for such purposes; it calculates QCD matrix elements by computing the QCD path integral numerically using Monte Carlo methods on a discretized spacetime grid. When the external states involve two or more nucleons, lattice QCD calculations become unviable; instead, one resorts to nuclear many-body methods. They involve two basic ingredients: (1) Highly precise few-body nuclear forces inferred from experimental data, and (2) special computational techniques to solve the many-body Schr\"{o}dinger equation to obtain nuclear states. For (2), the last two decades have seen a drastic progress in the developments of so-called \textit{ab initio} nuclear many-body methods; examples include No-Core Shell Model (NCSM)~\cite{Barrett:2013nh}, Coupled-Cluster (CC)~\cite{Hagen:2013nca} and In-Medium Similarity Renormalization Group (IMSRG)~\cite{Tsukiyama:2010rj,Hergert:2015awm} that are based on shell-model constructions, as well as Quantum Monte Carlo (QMC)~\cite{Carlson:2014vla} and Nuclear Lattice Effective Field Theory (NLEFT)~\cite{Lahde:2019npb} that are based on Monte Carlo approaches similar to lattice QCD. The existence of different \textit{ab initio} methods to compute the same nuclear matrix element provides a consistency check of the robustness of the final result.
\end{enumerate}

Having all the necessary notations stated and the general strategy outlines, below let us provide a few case study for RC to weak interaction processes. 

\section{Charged weak decay: Lifetime}

Testing the unitarity of the CKM matrix at the first row:
\begin{equation}
|V_{ud}|^2+|V_{us}|^2+|V_{ub}|^2=1
\end{equation}
provides a powerful probe of new physics at multi-TeV scale~\cite{Cirigliano:2023nol}. Among the various matrix elements, $|V_{ud}|\sim 0.97$ is numerically the largest and therefore demands the highest precision in its extraction. It can be extracted from variuos CW decays: semileptonic pion decay, free neutron decay, and nuclear beta decays. The extraction is a combined effort from experiment and theory. Schematically, it looks like:
\begin{equation}
\frac{1}{\tau}=|V_{ud}|^2G_F^2\times|M_\text{had}|^2\times(1+\Delta_R)\times F_\text{kin}~,
\end{equation}
where $\tau$ is the partial lifetime, $G_F$ is Fermi's constant, $M_\text{had}$ is the tree-level hadronic matrix element, $F_\text{kin}$ is a kinematic factor coming from the phase space integral, which may also include leading EM effects such as the Fermi function~\cite{Fermi:1934hr} as well as recoil corrections, and $\Delta_R$ is the radiative correction. Experimentally, one needs to measure $\tau$ and the $Q$-value that enters $F_\text{kin}$; the part of $M_\text{had}$ that involves axial current may also be measured through decay correlations. Theoretically, one needs to compute $M_\text{had}$ (if it is not protected by symmetry), the SM corrections to $F_\text{kin}$, as well as $\Delta_R$. 

\begin{figure}[tb]
	\includegraphics[scale=0.15]{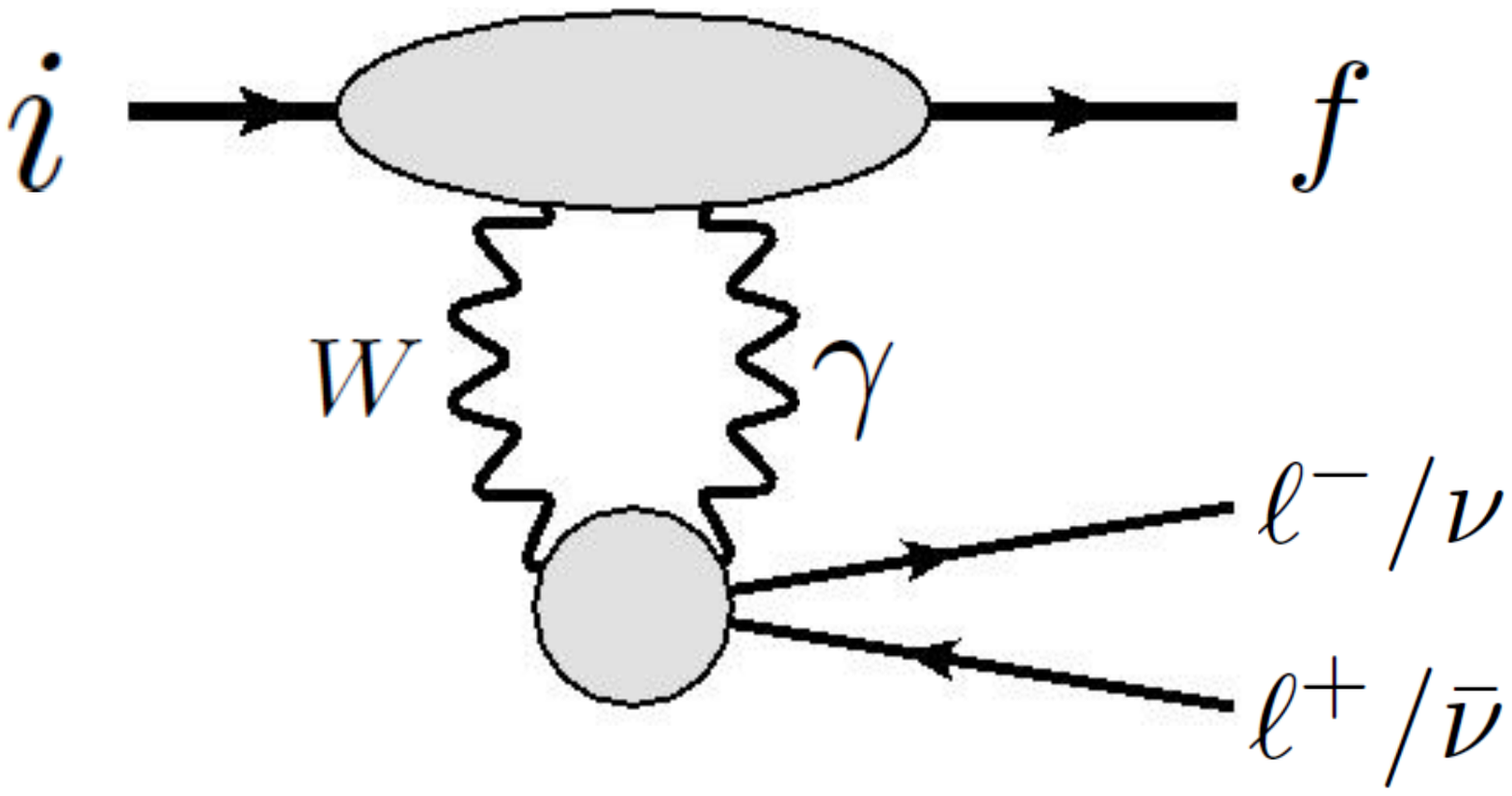}\hfill
	\caption{\label{fig:gammaW}The $\gamma W$-box diagram.}
\end{figure}

The leading contribution to $\Delta_R$, which scales as $\mathcal{O}(\alpha)$, consists of one-loop diagrams with a virtual photon and bremsstrahlung contributions with the emission of a single real photon. Diagrams where the photon only interact with the lepton are perturbatively calculable using standard field theory approaches, but those with photons interacting with the hadrons are not. So, the first task is to isolate all terms in the $\mathcal{O}(\alpha)$ RC that are not perturbatively calculable from the rest. The first comprehensive approach for this purpose was developed by Sirlin~\cite{Sirlin:1977sv}, which was based on rigorous Ward identities derived from the current algebra between electroweak currents; these identities are protected from QCD corrections. Details of the method can be found in Ref.\cite{Seng:2021syx}, and here we simply state the main conclusion: as far as the RC to the partial decay lifetime $\tau$ is concerned, the only non-perturbative contribution to the $\mathcal{O}(\alpha)$ RC resides in the so-called $\gamma W$-box diagram (see Fig.\ref{fig:gammaW}), where the lepton exchanges a photon and a $W$-boson simultaneously with the hadron. Neglecting recoil corrections, the upper shaded blob in this diagram is exactly the generalized Compton tensor $T^{\mu\nu}$ defined in Eq.\eqref{eq:Compton}, with $a=W$. The corresponding one-loop integral takes the following form:
\begin{equation}
\Box_{\gamma W}=\frac{ie^2}{2M^2}\int\frac{d^4q}{(2\pi)^4}\frac{M_W^2}{M_W^2-q^2}\frac{1}{q^2}\epsilon^{\mu\nu\alpha\beta}q_\alpha p_\beta \frac{T_{\mu\nu}}{F_+}\ ,\label{eq:BoxgammaW}
\end{equation}
where $M$ is the mass of the external hadron, $F_+$ is the tree-level decay amplitude, and the epsilon tensor comes from the lepton vertices and propagator. This integral picks up the anti-symmetric component of $T_{\mu\nu}$. In what follows, we will use free neutron decay as an illustration of how this quantity is computed non-perturbatively.

\section{Free neutron decay}

The free neutron decay $n\rightarrow p+e^-+\bar{\nu}_e$ is one of the most well-studied beta decay processes. At tree level, neglecting recoil corrections, the $W$-boson couples with the nucleon through the following weak current matrix element:
\begin{equation}
\langle p|J_W^\mu |n\rangle =\bar{u}_p\gamma^\mu(\mathring{g}_V+\mathring{g}_A\gamma_5)u_n\ ,
\end{equation}
where $\mathring{g}_V$ and $\mathring{g}_A$ are the (pure-QCD) vector and axial coupling constants, respectively. The vector coupling constant is protected by isospin symmetry, which gives $\mathring{g}_V=1$, while the axial coupling constant is not protected: $\mathring{g}_A\approx-1.27$. RC  renormalize the two couplings as $\mathring{g}_V\rightarrow g_V$, $\mathring{g}_A\rightarrow g_A$, and the ratio of the two normalized couplings, $\lambda\equiv g_A/g_V$, can be directly measured from various neutron decay correlations (note that we define $\lambda<0$ throughout this paper).

Neutron decay provides one of the primary avenue for the extraction of $|V_{ud}|$. This is done through the following master formula:
\begin{equation}
|V_{ud}|_n^2=\frac{5024.7~\text{s}}{\tau_n(1+3\lambda^2)(1+\Delta_R^V)}~.\label{eq:Vudn}
\end{equation}
In the expression above, the experimental inputs are the decay lifetime $\tau_n$ as well as the axial-to-vector ratio $\lambda$. The $5024.7\text{ s}$ in the numerator arises from the phase space integration, corrected by the Fermi function, recoil corrections and the energy-dependent part of the EM RC, which can be computed assuming nucleons are point-like particles. The remaining factor $1+\Delta_V^R$ in the denominator encodes the universal, ultraviolet piece of the electroweak RC, as well as the hadron-structure-dependent EM RC.

The box diagram integral in Eq.\eqref{eq:BoxgammaW} contributes to the renormalization of the vector and axial coupling constant:
\begin{equation}
g_V=\mathring{g}_V(1+\Box_{\gamma W}^V+\dots)\ ,\ g_A=\mathring{g}_A(1+\Box_{\gamma W}^A+\dots)\ . 
\end{equation}
Utilizing the decomposition of $T^{\mu\nu}$ in Eq.\eqref{eq:Tmunudecompose}, one obtains the following formula:
\begin{eqnarray}
\Box_{\gamma W}^V &=&\frac{e^2}{2m_N \mathring{g}_V}\int\frac{d^4q}{(2\pi)^4}\frac{M_W^2}{M_W^2+Q^2}\frac{1}{(Q^2)^2}\frac{\nu^2+Q^2}{\nu}T_3(\nu,Q^2)\nonumber\\
\Box_{\gamma W}^A &=&\frac{e^2}{m_N \mathring{g}_A}\int\frac{d^4q}{(2\pi)^4}\frac{M_W^2}{M_W^2+Q^2}\frac{1}{(Q^2)^2}\left\{\frac{\nu^2-2Q^2}{3\nu}S_1(\nu,Q^2)-\frac{Q^2}{\nu}S_2(\nu,Q^2)\right\}\label{eq:BoxCA}
\end{eqnarray}
In particular, the first line contributes to $\Delta_R^V$:
\begin{equation}
\Delta_R^V=2\Box_{\gamma W}^V+\dots\ ,
\end{equation}
where ``$+\dots$'' are all remaining terms that depend only on physics at the weak scale and can be calculated perturbatively.

\subsection{Dispersive analysis\label{sec:DRneutron}}

Below we introduce the dispersive treatment of the integrals in Eq.\eqref{eq:BoxCA}, interested readers can refer to Ref.\cite{Gorchtein:2023srs} for more details. It is apparent that the integrals depend only onf the $\nu$-odd component of the invariant amplitudes $T_3$, $S_1$ and $S_2$; one can show using isospin symmetry that this component arises from the isosinglet piece of the EM current, so we may denote it with a superscript (0). They can be connected to the structure functions of in Eq.\eqref{eq:Wmunu} through an unsubtracted dispersion relation:
\begin{eqnarray}
	T_3^{(0)}(\nu,Q^2)&=&-4i\nu\int_0^\infty d\nu'\frac{F_3^{(0)}(\nu',Q^2)}{\nu^{\prime 2}-\nu^2}\nonumber\\
	S_1^{(0)}(\nu,Q^2)&=&-4i\nu\int_0^\infty d\nu'\frac{g_1^{(0)}(\nu',Q^2)}{\nu^{\prime 2}-\nu^2}\nonumber\\
	S_2^{(0)}(\nu,Q^2)&=&-4i\nu\int_0^\infty\frac{g_2^{(0)}(\nu',Q^2)}{\nu^{\prime 2}-\nu^2}=-4i\nu^3\int_0^\infty\frac{g_2^{(0)}(\nu',Q^2)}{\nu^{\prime 2}(\nu^{\prime 2}-\nu^2)}\ .\label{eq:DRT3S1S2}
\end{eqnarray}
Note that in the last line we have further simplified the DR of $S_2^{(0)}$ using the Burkhardt-Cottingham sum rule~\cite{Burkhardt:1970ti}. Substituting them into Eq.\eqref{eq:BoxCA} gives the dispersive representation:
\begin{eqnarray}
	\Box_{\gamma W}^V&=&\frac{\alpha}{\pi \mathring{g}_V}\int_0^\infty \frac{dQ^2}{Q^2}\frac{M_W^2}{M_W^2+Q^2}\int_{0}^{1}dx\frac{1+2r}{(1+r)^2}F_3^{(0)}(x,Q^2)\nonumber\\
	\Box_{\gamma W}^A&=&-\frac{2\alpha}{\pi \mathring{g}_A}\int_0^\infty \frac{dQ^2}{Q^2}\frac{M_W^2}{M_W^2+Q^2}\int_0^1\frac{dx}{(1+r)^2}\left\{\frac{5+4r}{3}g_1^{(0)}(x,Q^2)-\frac{4m_N^2x^2}{Q^2}g_2^{(0)}(x,Q^2)\right\}\ ,
\end{eqnarray}
where we have defined the Bjorken variable $x=Q^2/(2m_N\nu)$, and $r=\sqrt{1+4m_N^2x^2/Q^2}$. Among the structure functions, $F_3^{(0)}$ and $g_1^{(0)}$ contribute in the UV regime ($Q^2\rightarrow \infty$). Such contribution can be computed using perturbative QCD (pQCD):
\begin{equation}
	\int_0^1 dx F_3^{(0)}(x,Q^2)=\frac{\mathring{g}_V}{6}C_\text{Bj}(Q^2)\ ,\ \int_{0}^{1}dx g_1^{(0)}(x,Q^2)=\frac{|\mathring{g}_A|}{12}C_\text{Bj}(Q^2)\ ,\label{eq:sumrule}
\end{equation}
where $C_\text{Bj}(Q^2)$ is the pQCD correction to the polarized Bjorken sum rule, which is currently known to $\mathcal{O}(\alpha_s^4)$~\cite{Baikov:2010iw,Baikov:2010je,Baikov:2012zn}.

Now we discuss the data inputs used to compute the dispersive integrals. The main challenge is that the structure functions above are all flavor non-diagonal, so it cannot be directly mapped to experimental cross sections without further manipulations. For the spin-dependent structure functions, isospin symmetry relates the isosinglet component $g^{(0)}$ to the corresponding EM structure functions:
\begin{equation}
g^{(0)}_{1,2}=\frac{1}{2}\left\{g_{1,2}^p-g_{1,2}^n\right\}\ ,\label{eq:g1g2}
\end{equation}
which means one can directly apply experimental data from EM interactions. On the other hand, $F_3^{(0)}$ cannot be related to EM structure functions because it is parity-odd. In principle, it can be related to the structure function $F_3$ measured in parity-violating deep inelastic scattering (PVDIS) that involves the interference between the EM and NW current,
but data of the latter are not currently available. Therefore, data analysis in $\Box_{\gamma W}^V$ must involve additional layers of theory modelings and thus is more complicated.

We start from the Born (elastic intermediate state) contribution to the structure functions, which is given by:
\begin{eqnarray}
F_3^{(0),B}(x,Q^2)&=&-\frac{1}{2}G_A(Q^2)G_M^S(Q^2)\delta(1-x)\nonumber\\
g_1^{(0),B}(x,Q^2)&=&\frac{F_1^V(Q^2) G_M^S(Q^2)+F_1^S(Q^2) G_M^V(Q^2)}{8}\delta(1-x)\nonumber\\
g_2^{(0),B}(x,Q^2)&=&-\tau\frac{F_2^V(Q^2) G_M^S(Q^2)+F_2^S(Q^2)G_M^V(Q^2)}{8}\delta(1-x)\ .\label{eq:elastic}
\end{eqnarray}
The expressions above are given entirely in terms of nucleon form factors: $F_{1,2}^V=F_{1,2}^p-F_{1,2}^n$, $F_{1,2}^S=F_{1,2}^p+F_{1,2}^n$, $G_M^V=G_M^p-G_M^n$, $G_M^S=G_M^p+G_M^n$ involve the proton and neutron EM form factors, while $G_A(Q^2)$ is the nucleon axial form factor, with $\tau=Q^2/(4m_N^2)$. The large amount of nucleon form factor data~\cite{Bernard:2001rs,Bhattacharya:2011ah,Lorenz:2012tm,Lorenz:2014yda,Ye:2017gyb,Lin:2021umk,Lin:2021umz,Borah:2020gte,Tomalak:2026wsu} allows us to pin down this contribution to high precision. 

Next we turn to non-elastic contributions. For $g_{1,2}^{(0)}$, Eq.\eqref{eq:g1g2} allows us to make use of data from single-nucleon polarized DIS experiments. Measurements of the structure function $g_1^N$ were carried out in SLAC~\cite{Anthony:1993uf,Abe:1994cp,Abe:1997cx}, CERN~\cite{Adams:1994zd,Alexakhin:2006oza,Alekseev:2010hc,Aghasyan:2017vck}, DESY~\cite{Ackerstaff:1997ws} and JLab~\cite{Deur:2004ti,Wesselmann:2006mw,Deur:2008ej,Guler:2015hsw,Fersch:2017qrq}.  Ref.\cite{Gorchtein:2021fce} made use of the results from the EG1b experiment at JLab that measured the $g_1^p$~\cite{Fersch:2017qrq} and $g_1^n$~\cite{Guler:2015hsw} in a wide range of $\{x,Q^2\}$, which returns a highly precise determination of $\Box_{\gamma W}^A$:
\begin{equation}
\Box_{\gamma W}^A=3.96(6)\times 10^{-3}\ .
\end{equation}

\begin{figure}[tb]
	\includegraphics[scale=0.3]{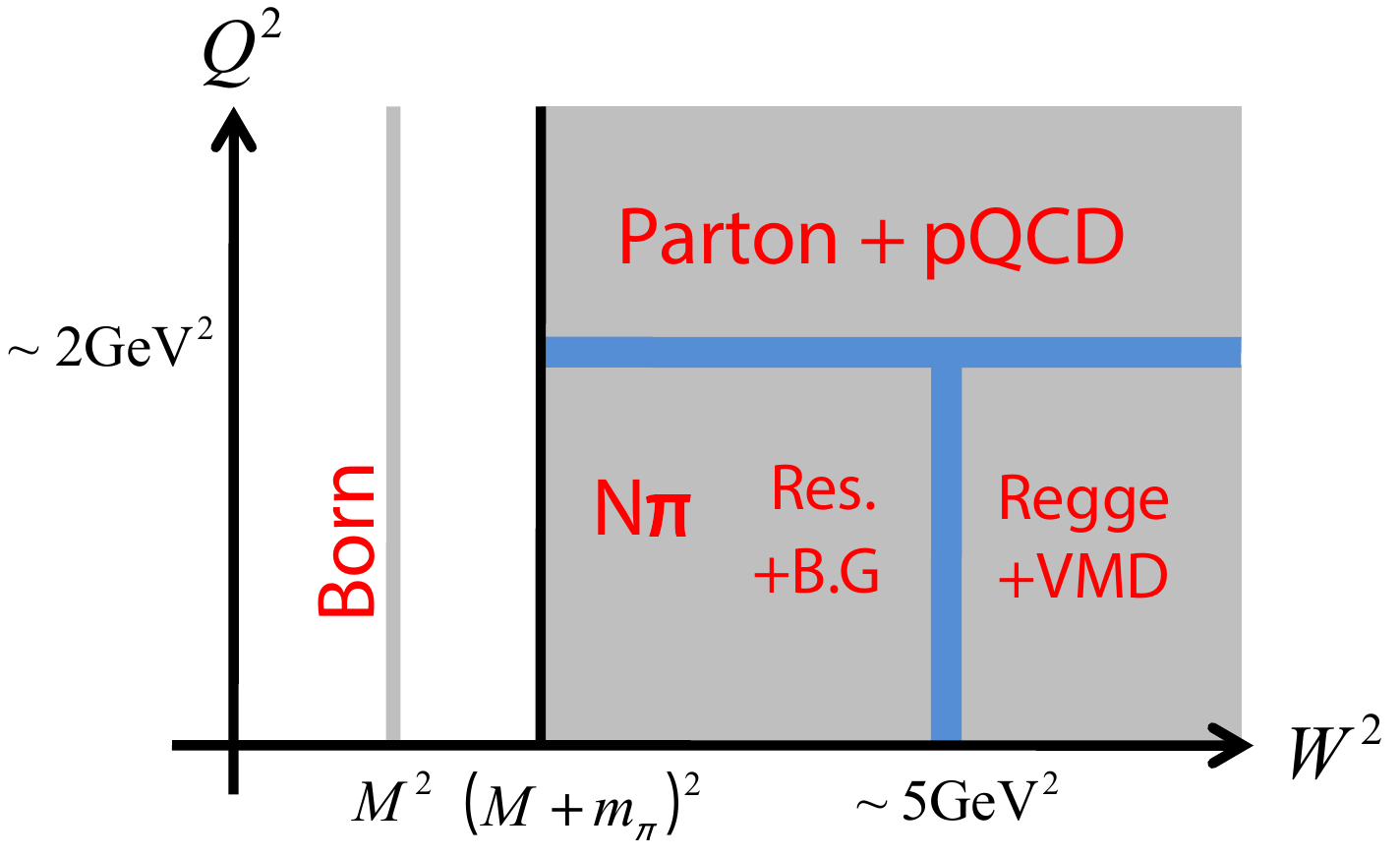}\hfill
	\caption{\label{fig:W-Q2diag} Approximate kinematical regions dominated by various physical mechanisms that contribute to $F_3^{(0)}$.}
\end{figure}

On the other hand, the structure function $F_3^{(0)}$ involves an axial current and cannot be matched into pure EM-induced inclusive scattering data. Furthermore, as stated above, inclusive electron-nucleon PVDIS scattering data at all kinematical regions is not yet available, so we need to resort to other weak scattering data, but now in a intermediate-state-dependent way. 
The way to proceed is to first divide the kinematical space of the structure functions into different regions of $Q^2$ and $W^2=(p+q)^2$, see Fig.\ref{fig:W-Q2diag}. The inelastic contributions start from the pion production threshold, $W_\pi^2=(m_N+m_\pi)^2$. In the inelastic region, when the virtuality of the gauge bosons is large enough  (roughly, $Q^2>2\text{ GeV}^2$), QCD becomes perturbative and one needs only the pQCD-corrected sum rules of structure functions, as given in Eq.\eqref{eq:sumrule}. At small $Q^2$ and $W^2$, the main contributors are $N\pi$ and nucleon resonances. The former can be computed using low-energy effective theories such as chiral perturbation theory, while for each resonance one can compute it contribution using known resonance transition form factors~\cite{Lalakulich:2005cs,Lalakulich:2006sw}, much like the case of elastic form factors~\eqref{eq:elastic}. Numerically, the contribution in this region is shown to be negligibly small (due to the fact that $\Delta$ does not contribute) so the theory uncertainty is not an issue.

\begin{figure}[tb]
	\includegraphics[scale=0.2]{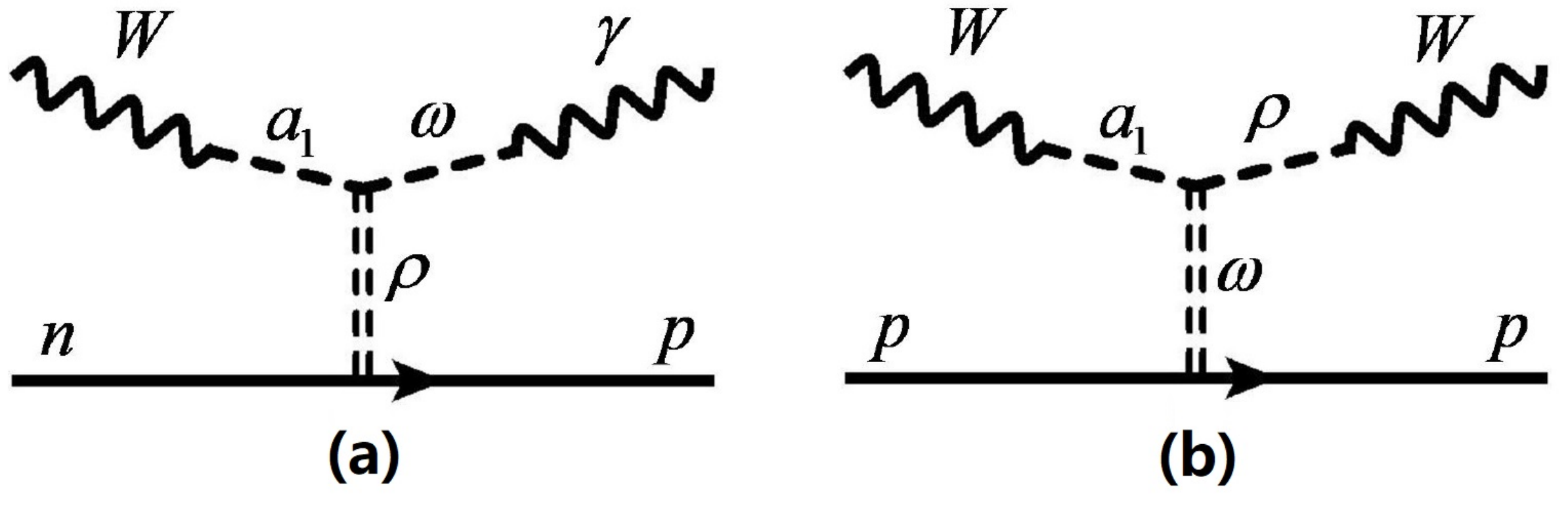}\hfill
	\caption{\label{fig:Regge} Diagrammatic illustration of the Regge exchange picture description for $F_3^{(0)}$ (left) and $F_3^{W}$ (right).}
\end{figure}

The real challenge comes from the large-$W^2$, small-$Q^2$ region, where the structure functions are contributed by multi-hadron intermediate states (which effect also penetrates to the small-$W^2$ region as a smooth background). It is not possible to study the impact from each of the intermediate states separately. Fortunately, there is an economical way, known as the ``Regge exchange picture'', to describe the high-energy contribution collectively. In this picture, the two virtual gauge bosons first fluctuate into a vector or axial meson, which then interacts with the nucleon through the exchange of a ``Regge trajectory'', which resembles yet another vector or axial meson (in terms of quantum numbers). This is depicted in Fig.\ref{fig:Regge}; for the case of $F_3^{(0)}$ (left diagram), the photon and $W$-boson fluctuate into a $\omega$ and $a_1$ respectively, which then exchange a $\rho$-trajectory with the nucleon.  

The advantage of the picture is that it offers a natural connection between the structure function in $\Box_{\gamma W}$ and that in other weak processes. An example is the inclusive neutrino/antineutrino-proton charged-current scattering $\nu(\bar \nu) +p\rightarrow e^{\mp}+X$ triggered by a $W$-exchange. The corresponding parity-odd structure function $F_3^{\nu p+\bar{\nu}p}$ results from the interference between the vector and the axial component of $J_W^\mu$. The corresponding Regge exchange picture is given in the right diagram in Fig.\ref{fig:Regge}: the two $W$-bosons fluctuate into a $\rho$ and a $a_1$ respectively, which then exchanges an $\omega$ trajectory with the proton. An important observation is that the three-meson coupling in the two diagrams of Fig.\eqref{fig:Regge} are identical (and that propagators of the $\rho$ and $\omega$ trajectories are almost identical due to their near mass degeneracy), which suggests that the $\{W^2,Q^2\}$-dependence of the Regge contribution ($\mathbb{R}$) in these two cases are the same, so to map $F_3^{(0)}$ to $F_3^{\nu p+\bar{\nu}p}$ one needs only to compute a constant rescale factor that accounts for the difference in the gauge boson-meson fluctuation and the meson-nucleon coupling constants. That amounts to:
\begin{equation}
	F_{3,\mathbb{R}}^{(0)}(x,Q^2)=\frac{1}{18}F_{3,\mathbb{R}}^{\nu p+\bar{\nu}p}(x,Q^2)\ ,\label{eq:Regge}
\end{equation}
where $F_3^{\nu p+\bar{\nu}p}=(F_3^{\nu p}+F_3^{\bar{\nu}p})/2$, with
	\begin{eqnarray}
	\frac{1}{4\pi}\sum_{X}(2\pi)^4\delta^4(p+q-p_X)\langle p|(J_W^\mu)^\dagger|X\rangle\langle X|J_W^\nu|p\rangle
	&=&-i\varepsilon^{\mu\nu\alpha\beta}\frac{q_\alpha p_\beta}{2p\cdot q}F_3^{\nu p}(\nu,Q^2)+\dots\nonumber\\
	\frac{1}{4\pi}\sum_{X}(2\pi)^4\delta^4(p+q-p_X)\langle p|J_W^\mu|X\rangle\langle X|(J_W^\nu)^\dagger|p\rangle
	&=&-i\varepsilon^{\mu\nu\alpha\beta}\frac{q_\alpha p_\beta}{2p\cdot q}F_3^{\bar{\nu} p}(\nu,Q^2)+\dots
\end{eqnarray}
Data is available for the structure function $F_3^{\nu p+\bar{\nu}p}$ obtained from neutrino scattering of light nuclei~\cite{Bolognese:1982zd,Aachen-Bonn-CERN-Democritos-London-Oxford-Saclay:1982gpl}, which can serve as the data input to $\Box_{\gamma W}^V$ through Eq.\eqref{eq:Regge}. Two data-driven calculations of $\Box_{\gamma W}^V$ based on the strategy approach was performed Refs.\cite{Seng:2018yzq,Seng:2018qru} utilized the aforementioned neutrino scattering data with integrated $x$, whereas Ref.\cite{Shiells:2020fqp} made use of the same set of data, but fit their functional form with both $x$ and $Q^2$-dependence.  

The first dispersive analysis of $\Box_{\gamma W}^V$ in 2018~\cite{Seng:2018yzq} had a profound impact on the status of the first-row CKM unitarity. It substantially increased the central value of $\Delta_R^V$ with reduced theory uncertainty: 
\begin{eqnarray}
\Delta_R^V&:&0.02361(38)\ ,\ \text{pre-2018}\nonumber\\
&\rightarrow&0.02467(22)\ ,\ \text{2018}
\end{eqnarray}
which subsequently reduced the central value of $|V_{ud}|$:
\begin{eqnarray}
|V_{ud}|&:&0.97420(38)\ ,\ \text{pre-2018}\nonumber\\
&\rightarrow&0.97366(15)\ .\ \text{2018}
\end{eqnarray}
This changed the status of the first-row CKM unitarity from almost perfectly satisfied to having a $\sim 4\sigma$ deviation (which significance was later slightly reduced due to increased nuclear uncertainty, as will be discussed later).

\subsection{Lattice QCD analysis}

Lattice QCD provides a completely independent and complementary non-perturbative approach to tackle the  same problem. On a discrete lattice of Euclidean spacetime, one computes the generalized Compton tensor $T^{\mu\nu}$ directly as a ``four-point correlation function'': two points from the external hadron states and two points from the current insertions. In terms of evaluating $\Box_{\gamma W}^V$, one needs the following $x$-integral of the structure function $F_3^{(0)}(x,Q^2)$, which is called its ``first Nachtmann moment''~\cite{Nachtmann:1973mr,Nachtmann:1974aj}:  
\begin{equation}
M_3^{(0)}(Q^2)\equiv \frac{4}{3}\int_0^1dx\frac{1+2r}{(1+r)^2}F_3^{(0)}(x,Q^2)\ .
\end{equation}
Notice that it reduces to a simple integration of $F_3^{(0)}(x,Q^2)$ over $x$ at larger $Q^2$. Lattice QCD is able to predict $M_3^{(0)}(Q^2)$ to high accuracy; however, as $Q^2$ increases, the result starts to suffer from large 
lattice artifacts (i.e. systematic uncertainties due to finite lattice spacing) and becomes unreliable; on the other hand, the pQCD computation (Eq.\eqref{eq:sumrule}) is reliable at large $Q^2$ but not at small $Q^2$. So, a fully first-principles treatment is achieved by combining the lattice result at low $Q^2$ and pQCD result at large $Q^2$. In doing so, one observes a smooth transition between these descriptions around $Q^2=2\text{ GeV}$, which justifies the region separation in Fig.\ref{fig:W-Q2diag}.

\begin{figure}[tb]
	\includegraphics[scale=0.3]{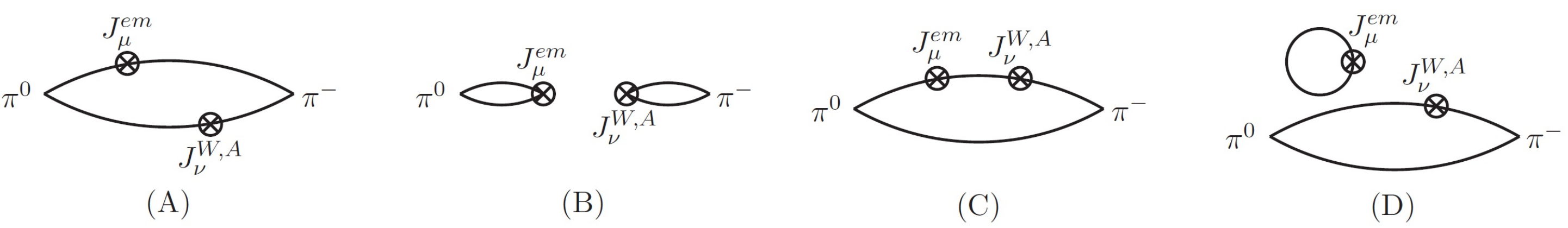}\hfill
	\caption{\label{fig:contraction} Contraction diagrams involved in the lattice calculation of $T^{\mu\nu}$ for semileptonic pion decay.}
\end{figure}

The first lattice computation of $\gamma W$-box diagram was performed in Ref.\cite{Feng:2020zdc} for semileptonic pion decay, $\pi^+\rightarrow \pi^0+e^++\nu_e$, by computing the contraction diagrams in Fig.\ref{fig:contraction}. The result was later independently confirmed in Ref.\cite{Yoo:2023gln}. It achieved a 3-times reduction of the hadronic uncertainty in the radiative corrections compared to the previous state-of-the-art result based on chiral perturbation theory~\cite{Cirigliano:2002ng}, which effectively turns this decay to the theoretical cleanest channel to measure $|V_{ud}|$~\cite{Cirigliano:2026ios}. In addition, the pion box diagram result can be used to infer the Regge contribution to the neutron box diagram in the dispersive approach~\cite{Seng:2020wjq}.
A complete lattice QCD calculation of $\Box_{\gamma W}^{V}$ and $\Box_{\gamma W}^{A}$ in free neutron decay was first accomplished in Ref.\cite{Ma:2023kfr}.

\section{Box diagram in the nuclear level}

The uncertainties in the extraction of $|V_{ud}|$ from both the semileptonic pion decay and free decay are both dominated by experimental errors. For semileptonic decay, it is from the semileptonic decay branching ratio, and for neutron decay it is from both the neutron lifetime and the coupling constant ratio $\lambda$. The situation is a bit different for nuclear beta decays. An important avenue for the extraction of $|V_{ud}|$ is the ``superallowed'' $0^+\rightarrow 0^+$ nuclear beta decays; it consists of 23 measured transitions from ${}^{10}\text{C}\rightarrow{}^{10}\text{B}$ to ${}^{74}\text{Rb}\rightarrow{}^{74}\text{Kr}$, 15 of which lifetime precision is better than 023\%~\cite{Hardy:2020qwl}. Averaging over all such transitions effectively reduces the experimental uncertainty, with the price of introducing new nuclear-structure-dependent theory uncertainties not present in pion and neutron. At the present, the $|V_{ud}|$ extracted from superallowed nuclear beta decays is claimed to have the highest precision, which imposes strong constraints on new physics~\cite{Brodeur:2023eul}.      

The master formula for the $V_{ud}$ extraction from superallowed beta decays reads~\cite{Hardy:2020qwl}:
\begin{equation}
	|V_{ud}|^2_{0^+}=\frac{2984.43~\text{s}}{ft(1+\delta_R')(1+\delta_\text{NS}-\delta_\text{C})(1+\Delta_R^V)}\ .
\end{equation}
Among the quantities at the right hand side, $\delta_R'$, $\Delta_R^V$ and $\delta_\text{NS}$  originate from radiative corrections: $\delta_R'$ is the so-called ``outer'' radiative correction that it structure independent~\cite{Sirlin:1967zza,Cao:2025zxs,Crosas:2025xyv}; $\Delta_R^V$ is exactly the same single-nucleon RC that appears in the free neutron formula, Eq.\eqref{eq:Vudn}; finally, $\delta_\text{NS}$ is the nuclear-structure (NS)-dependent part of the radiative correction. The separation of the full structure-dependent correction into $\Delta_R^V$ and $\delta_\text{NS}$ is just a convention; they both arise from the nuclear $\gamma W$-box diagrams, where the $i$ and $f$ in Fig.\ref{fig:gammaW} are the nuclear external states. 

Computing $\delta_\text{NS}$ requires the handling of many-nucleon contributions to $T^{\mu\nu}$, which 
calls for nuclear many-body methods. There are currently two approaches to do so; we will describe the details here, and interested readers may refer to  Refs.\cite{Gorchtein:2023naa,Sargsyan:2026ygi} for a more comprehensive review.
\begin{enumerate}
	\item \textbf{Current algebra approach}: In this approach developed in Refs.\cite{Seng:2018qru,Seng:2022cnq}, one simply applie Eq.\eqref{eq:BoxCA} at the nuclear level, and compute the invariant amplitude $T_3$ using many-body methods. A useful representation is\footnote{The normalization of $T_3$ depends on that of $T^{\mu\nu}$, which can be different in different literature.}: 
	\begin{equation}
		T_3(\nu,Q^2)=-\frac{4M\nu}{|\vec{q}|}\sum_X\Bigg[\frac{\langle \phi_f|J_\text{em}^x(\vec{q})|X\rangle\langle X|J_{W5}^{\dagger y}(-\vec{q})|\phi_i\rangle}{\nu_X-\nu-i\varepsilon}
		+\frac{\langle \phi_f|J_{W5}^{\dagger y}(-\vec{q})|X\rangle\langle X|J_\text{em}^x(\vec{q})|\phi_i\rangle}{\nu_X+\nu-i\varepsilon}\Bigg]~,
		\label{eq:T3GF}
	\end{equation}
	where $X$ sums over all intermediate nuclear states. In other words, one needs to compute a nuclear Green's function: $G(z)=1/(z-H)$, where $H$ is the nuclear Hamiltonian.  
	\item \textbf{Effective field theory (EFT) approach}: In this approach developed in Refs.\cite{Cirigliano:2024msg,Cirigliano:2024rfk}, instead of computing the nuclear Green's function, one constructs effective two-nucleon potentials that encodes the RC. It corrects the CW current as:
	\begin{eqnarray}
		\mathcal{J}_W^\mu&=&\sum_{n=1}^A(g_V \delta^{\mu 0}-|g_A|\delta^{\mu i}\sigma^{(n)i})\tau^{(n)+}+(\mathcal{J}^{2b})^\mu +\dots\nonumber\\
		&&+\delta^{\mu 0}(\mathcal{V}^0+E_0\mathcal{V}_E^0)+\delta^{\mu i}\mathcal{V}_i+p_e^\mu\mathcal{V}_{m_e}+\dots~,\label{eq:VE}\ ,
	\end{eqnarray}
where the effective potentials $\mathcal{V}^0$, $\mathcal{V}_E^0$, $\mathcal{V}_i$ and $\mathcal{V}){m_e}$ are derived from two-nucleon Feynman diagrams based on chiral effective field theory. The most piece among them is the energy-independent potential $\mathcal{V}^0$, which consists of a long-distance and a short-distance piece, the latter shows up as a contact interaction with unknown low energy constants (LECs) that incorporate short-distance effects not captured by chiral symmetry.  
\end{enumerate}

Both methods have their own advantages and disadvantages. The current algebra approach is more computationally demanding because it involves a summation over all intermediate states, but doing so explicitly includes all nuclear effects. The EFT approach, on the other hand, is simpler because it requires only the computation of ground state nuclear matrix elements of the relevant two-nucleon potentials; the price to pay is the existence of unknown LECs that give a large theory uncertainty. At the present, the current algebra approach has been applied to ${}^{10}\text{C}\rightarrow {}^{10}\text{B}$~\cite{Gennari:2024sbn}, while the EFT approach has been applied to ${}^{14}\text{O}\rightarrow {}^{14}\text{N}$~\cite{Cirigliano:2024msg} and ${}^{10}\text{C}\rightarrow {}^{10}\text{B}$~\cite{King:2025fph}.

\section{Charged weak decay: Correlations}

In the above, we discussed the impact of radiative corrections to beta decay lifetime. However, there are much more observables beta decay processes that carry information of new physics. Examples are the ``decay correlations'',  namely the dependence of the differential decay rate on the momentum or spin of the initial particle or (detectable) decay products. Let us take the free neutron decay as an example: At tree level, the differential decay rate possesses the following correlation structures:
\begin{equation}
	d\Gamma\propto 1+a\frac{\vec{p}_e\cdot\vec{p}_\nu}{E_eE_\nu}+b\frac{m_e}{E_e}+\hat{s}_n\cdot\left[A\frac{\vec{p}_e}{E_e}+B\frac{\vec{p}_\nu}{E_\nu}+\dots\right]\ ,\label{eq:correlations}
\end{equation}
where $\hat{s}_n$ is the neutron unit spin vector. The constants $a$, $b$, $A$, $B$ and so on are called ``correlation coefficients''. In the SM, they are given by:
\begin{equation}
a=\frac{1-\lambda^2}{1+3\lambda^2}\ ,\ b=0\ ,\ A=-2\lambda\frac{\lambda+1}{1+3\lambda^2}\ , B=2\frac{\lambda-1}{1+3\lambda^2}\ .
\end{equation}
An immediate application of these relations is the  measurement of the quantity $\lambda$ from various correlations, and all the outcomes should be identical if SM is correct. A discrepancy in the $\lambda$-determination from different correlation may be a hint of new physics. This turns out to be the case at the present,  as the best determinations of $\lambda$ from $A$~\cite{Markisch:2018ndu} and $a$~\cite{Beck:2019xye,Beck:2023hnt} exhibit a discrepancy of about $3.5\:\sigma$. 

To extract $\lambda$ at high precision, one needs to account for the RC to the correlation coefficients. Hadron-structure-dependent RCs, such as $\Box_{\gamma W}^V$ and $\Box_{\gamma W}^A$ in Eq.\eqref{eq:BoxCA}, are reabsorbed into the measured value of $\lambda$ and hence does not alter the correlation structure in Eq.\eqref{eq:correlations}. Instead, we are interested in the part of the RC that depends on the lepton energies, which can be computed using elementary QED by treating the nucleon as a point particle, with the effective vector and axial coupling constants $g_V$ and $g_A$. This was first performed in Refs.\cite{Sirlin:1967zza,Garcia:1981it}, where the RC-corrected differential decay rate was expressed in terms of the electron momentum $\vec{p}_e$ and the neutrino solid angle $\Omega_\nu$. Their result reads:
\begin{eqnarray}
	\frac{d\Gamma}{dE_ed\Omega_ed\Omega_\nu}&=&\frac{(G_F V_{ud})^2}{(2\pi)^5}|\vec{p}_e|E_e(E_m-E_e)^2F(E_e)g_V^2(1+3\lambda^2)\left(1+\frac{\alpha}{2\pi}\delta^{(1)}(E_e)\right)\nonumber\\
	&&\times\left\{1+\left(1+\frac{\alpha}{2\pi}\delta^{(2)}(E_e)\right)a\frac{\vec{p}_e\cdot\vec{p}_\nu}{E_eE_\nu}+\hat{s}_n\cdot\left[\left(1+\frac{\alpha}{2\pi}\delta^{(2)}(E_e)\right)A\frac{\vec{p}_e}{E_e}+B\frac{\vec{p}_\nu}{E_\nu}+\dots\right]\right\}\ ,\label{eq:outerRCold}
\end{eqnarray}
where
\begin{eqnarray}
\delta^{(1)}(E_e)&=&3\ln\frac{m_p}{m_e}-\frac{3}{4}+4\left(\frac{1}{\beta}\tanh^{-1}\beta-1\right)\left(\ln\frac{2(E_m-E_e)}{m_e}+\frac{E_m-E_e}{3E_e}-\frac{3}{2}\right)-\frac{4}{\beta}\text{Li}_2\left(\frac{2\beta}{1+\beta}\right)\nonumber\\
&&+\frac{1}{\beta}\tanh^{-1}\beta\left(2+2\beta^2+\frac{(E_m-E_e)^2}{6E_e^2}-4\tanh^{-1}\beta\right)\nonumber\\
\delta^{(2)}(E_e)&=&2\left(\frac{1-\beta^2}{\beta}\right)\tanh^{-1}\beta+\frac{4(E_m-E_e)(1-\beta^2)}{3\beta^2E_e}\left(\frac{1}{\beta}\tanh^{-1}\beta-1\right)+\frac{(E_m-E_e)^2}{6\beta^2E_e^2}\left(\frac{1-\beta^2}{\beta}\tanh^{-1}\beta-1\right)\label{eq:delta12}
\end{eqnarray}
where $\beta\equiv |\vec{p}_e|/E_e$, $E_m=(m_n^2-m_p^2+m_e^2)/(2m_n)$ is the electron's end-point energy, and $F(E)$ is the Fermi function that accounts for the Coulomb interaction between the proton and the outgoing electron~\cite{Fermi:1934hr}.
The result above includes both the one-loop corrections and bremsstrahlung, to ensure the cancellation of infrared divergence required by the Kinoshita-Lee-Nauenberg (KLN) theorem~\cite{Kinoshita:1962ur,Lee:1964is}. Notice that $\delta^{(1)}(E_e)$ includes a large logarithm $3\ln(m_p/m_e)$, but this is just a matter of choice since any constant term in $\delta^{(1)}(E_e)$ can be re-absorbed into the definition of the renormalized $g_V$; the latter is adopted in the similar calculation based on EFT~\cite{Cirigliano:2022hob}.

There is a hidden kinematic issue in the formalism above, namely it depends on the neutrino momentum $\vec{p}_\nu$ is not directly measured in experiment. One may think that it is deducible from the momentum conservation $\vec{p}_\nu=\vec{p}_n-\vec{p}_p-\vec{p}_e$, but this identity works only for three-body decay. At $\mathcal{O}(\alpha)$, one needs to include simultaneously the three-body and four-body decay, the latter including one real photon emission, to ensure IR finiteness. The inclusion of the four-body decay invalidates the identity above. In other words, Eq.\eqref{eq:outerRCold} is mathematically correct, but not directly applicable for experiments that do not observe the neutrino. If we are interested in correlation structures that do not depend on $\vec{p}_\nu$, for example $A$, one can integrate out $\Omega_\nu$ so the formula above is perfectly fine. However, for correlation structures that depend intrinsically on $\vec{p}_\nu$, for example $a$ and $B$, the formula above is not applicable in practice. 

The way to overcome this defect is simply to avoid using $\vec{p}_\nu$ as the variable in the differential decay rate formula. There are two ways to implement this: (1) To probe spin-independent correlations (such as $a$ and $b$), one can use the electron energy $E_e$ and the proton kinetic energy $K_p$ as independent variables; this is called the ``recoil formalism''~\cite{Gluck:2022ogz}. The second way, known as the ``pseudo-neutrino formalism'', is to define $\vec{p}_\nu\equiv \vec{p}_n-\vec{p}_e$, and use $\{\vec{p}_e,\Omega_\nu'\}$ (instead of $\{\vec{p}_e,\Omega_\nu\}$) as independent variables~\cite{Seng:2023ynd}. This formalism is useful in the sense that it resembles the traditional approach, and is also applicable to spin-dependent correlations such as $B$ that requires angular measurements.

\section{Neutral current interaction}
In the above, we discussed the search for new physics in CW decay processes; the same can be done in NW processes. The SM electroweak gauge group is $\text{SU(2)}_L\times \text{U(1)}_Y$, which gives two neutral gauge  bosons $W^0$ and $B$; these two gauge bosons ``mix'' to form the massless photon and the massive $Z$-boson. The mixing takes the following form:
\begin{equation}
\left(\begin{array}{c}
\gamma\\
Z
\end{array}\right)=\left(\begin{array}{cc}
c_{w} & s_{w}\\
-s_{w} & c_{w}
\end{array}\right)\left(\begin{array}{c}
B\\
W^{0}
\end{array}\right)\ ,
\end{equation}
which involves the weak mixing angle $\theta_W$ introduced in Eq.\eqref{eq:Lew}. At tree level, it is given by $s_w^2=1-M_W^2/M_Z^2$, but at higher order it depends on a renormalization scale $\mu$. The running of $s_w^2(\mu)$ (say,in the $\overline{\text{MS}}$ scheme) is well studied~\cite{Erler:2017knj}, which serves as a rigorous SM prediction that can be tested in experiments. Deviations from the SM prediction could indicate new physics, for example a dark $Z$-boson~\cite{Kumar:2013yoa,Davoudiasl:2015bua}.

As in Eq.\eqref{eq:Lew}, the neutral current depends on $s_w^2$, so by measuring the NW interaction strength one can extract $s_w^2$. The problem, however, is that the EM interaction strength is much larger than the NW strength, so ordinary experimental observables are dominated by EM contributions and are impossible to resolve the tiny NW contributions. One can make use of discrete symmetry to overcome this challenge. Given that EM interactions conserve parity while NW interactions do not, if we measure a parity-violating (PV) asymmetry, the leading contribution must be proportional to the NW interaction. In the following we discuss an example of such experiment, the parity-violating elastic scattering (PVES) between electron and nucleon/nucleus.

\subsection{Basics of nuclear weak charge}

Consider the scattering between and electron and an unpolarized, strongly-bounded state $\phi$, which can be a nucleon or a nucleus. At tree level, the interaction depends on two spin-independent matrix elements:
\begin{equation}
	\langle \phi(p')|J_\text{em}^\mu(0)|\phi(p)\rangle\equiv Q_\text{ch}^\phi F_\text{ch}(t)(p+p')^\mu~,~\langle \phi(p')|J_Z^\mu(0)|\phi(p)\rangle\equiv \frac{1}{2}Q_W^\phi F_W(t)(p+p')^\mu~,
\end{equation}  
where $t\equiv (p-p')^2$. $Q_\text{ch}^\phi$ and $Q_W^\phi$ are the electric and weak charge of $\phi$, respectively. The charge and weak form factors $F_\text{ch}(t)$ and $F_W(t)$ are both normalized to 1 at $t=0$. Since vector currents are conserved, the charges of a nucleus with $Z$ protons and $N$ neutron would simply be the sum of the respective charges of all nucleons:
\begin{equation}
	Q_\text{ch}^\phi=Z~,~Q_W^\phi=ZQ_W^p+NQ_W^n~,
\end{equation}
where the weak charges of the nucleon are, in turn, the sum of those from the quarks:
\begin{equation}
	Q_W^p=2(2g_V^u+g_V^d)=1-4s_w^2~,~Q_W^n=2(g_V^u+2g_V^d)=-1~.
\end{equation}
So, measuring the weak charge of a proton or nuclear target provides an access to the weak mixing angle. 

The PV asymmetry can be defined in terms of the difference between the differential scattering cross section of an electron with positive and negative helicity:
\begin{equation}
	A^{PV}\equiv \frac{d\sigma_+-d\sigma_-}{d\sigma_++d\sigma_-}~.
\end{equation}
At tree level, this gives:
\begin{equation}
A_\text{tree}^{PV}=\frac{G_Ft}{4\pi\sqrt{2}\alpha}\frac{Q_W^\phi F_W(t)}{ZF_\text{ch}t}~,
\end{equation}
from which the nuclear weak charge can be obtained in the forward limit:
\begin{equation}
	\left. Q_W^\phi\right|_\text{tree}=\left.\frac{4\pi\sqrt{2}\alpha}{G_Ft}ZA^{PV}\right|_{t\rightarrow 0}\ ,\label{eq:Qwtree}
\end{equation}
Taking $t\rightarrow 0$ is to avoid uncertainties from the form factors, but at the same time $A^{PV}$ also vanishes in the forward limit, so one must find a balance. 

A particularly interesting case is the proton target, because it just turns out that $s_w^2\approx 0.25$, so the proton weak charge $1-4s_w^2$ is accidentally suppressed. In this case, the SM contribution to $A^{PV}$ is suppressed so it is more sensitive to PV effects from new physics. Recently, the $Q_\text{weak}$ experiment at Jefferson Lab that performed an $ep$-PVES measured the PV asymmetry with a 6\% precision:
\begin{equation}
	A_\text{PV}=(-226.5\pm 9.5)\times 10^{-9}\ ,
\end{equation}
which implied the following value of the proton weak charge at $\mu\approx 0.1\text{ GeV}$:
\begin{equation}
	Q_W^p=0.0719\pm 0.0045\ ,
\end{equation} 
in excellent agreement with the SM prediction~\cite{Qweak:2018tjf}. In addition, the upcoming P2 experiment at Mainz plans to measure $A_{PV}$ with a 1.4\% precision, which will significantly improve the proton weak charge measurement~\cite{Becker:2018ggl}.

\subsection{Radiative corrections to $ep$-PVES}
  
Electroweak RC to the $ep$-PVES process modifies the value of the measured proton weak charge from the tree-level expression, Eq.\eqref{eq:Qwtree}. The $t\rightarrow 0$ limit removes the bremsstrahlung contribution because there is no phase space for the emission of an extra photon. This also implies that the loop contributions are free from infrared divergences. 
The RC-modified expression of the proton weak charge reads~\cite{Erler:2003yk}:
\begin{equation}
	Q_W^p=(1+\Delta\rho+\Delta_e)(1-4s_w^2+\Delta_e')+\Box_{WW}+\Box_{ZZ}+\Box_{\gamma Z}\ .
\end{equation}
Let us explain the meaning of each term. First, the factor $1+\Delta\rho$ renormalizes the ratio of the NW and CW current interaction strength at low energies~\cite{Veltman:1977kh}. Next, $\Delta_e$ and $\Delta_e'$ are corrections to the axial $Zee$ and $\gamma ee$ couplings, respectively~\cite{Marciano:1982mm,Marciano:1983ss}. The corrections $\Box_{WW}$ and $\Box_{ZZ}$ come from the $WW$ and $ZZ$-box diagrams, which depend only on physics at the weak scale and therefore can be computed perturbatively.

\begin{figure}[tb]
	\includegraphics[scale=1]{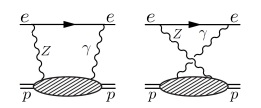}\hfill
	\caption{\label{fig:gammaZ}The $\gamma Z$-box diagram correction to $ep$ scattering.}
\end{figure}

The only one-loop diagram that involves non-perturbative QCD is the $\gamma Z$-box diagram depicted in Fig.\ref{fig:gammaZ}. The shaded blob in the figure is again the generalized forward Compton tensor $T^{\mu\nu}$ (that involves the neutral weak current, i.e. $a=Z$), which adopts the decomposition in Eq.\eqref{eq:Tmunudecompose}. The box diagram can be split into the ``vector'' (V) and ``axial'' (A) components, contributed by the vector and axial neutral weak current, respectively:
\begin{equation}
	\Box_{\gamma Z}=\Box_{\gamma Z}^V+\Box_{\gamma Z}^A\ .
\end{equation}
Following the notation in Ref.\cite{Zhang:2026utt}, they take the following form:
\begin{eqnarray}
	\Box_{\gamma Z}^V&=&\alpha\int\frac{d^4q}{(2\pi)^4}\sum_{i=1}^2 V_i(q,\ell)T_i^{\gamma Z}(\nu,\vec{q})\nonumber\\
	\Box_{\gamma W}^A&=&\alpha\int\frac{d^4q}{(2\pi)^4}V_3(q,\ell)T_3^{\gamma Z}(\nu,\vec{q})\label{eq:gammaZbox}
\end{eqnarray}
where 
\begin{eqnarray}
	V_1&=&-\frac{\ell\cdot q}{2q^2}(-2\ell\cdot q+3q^2)\nonumber\\
	V_2&=&\frac{C}{2(p\cdot q)(q^2)^2}\left[2(\ell\cdot q)^2(p\cdot q)^2+2(\ell\cdot p)^2(q^2)^2+(\ell\cdot q)q^2\left(-4(\ell\cdot p)(p\cdot q)-(p\cdot q)^2+m_N^2q^2\right)\right]\nonumber\\
	V_3&=&-\frac{1}{2}(1-4s_w^2)C\left(-\ell\cdot q+\frac{\ell\cdot p}{p\cdot q}q^2\right)
\end{eqnarray}
where
\begin{equation}
	C=\frac{16\pi}{\ell\cdot p}\frac{M_Z^2}{((\ell-q)^2+i\epsilon)(q^2+i\epsilon)(M_Z^2-q^2-i\epsilon)}\ ,\label{eq:propagators}
\end{equation}
with $\ell=(E,\vec{l})$ the momentum of the electron, where we have taken the $m_e\rightarrow 0$ limit for simplicity. Unlike nuclear beta decays where the lepton momentum is restricted by the phase space, here the energy of the incoming electron beam is fixed by the experimental setup and can be large; for instance, the $Q_\text{weak}$ experiment was conducted with $E=1.16\text{ GeV}$. Therefore, it is important to retain the $E$-dependence in the expressions above. 

\subsection{Dispersive analysis}

Apart from the known point-like proton intermediate state contribution, one can show that $\Box_{\gamma Z}^V$ vanishes when $E\rightarrow 0$ while $\Box_{\gamma Z}^A$ approaches a non-zero constant. So, earlier studies of $\Box_{\gamma Z}$ had always implicitly assumed $E=0$ and thus focused on the axial component~\cite{Marciano:1983ss}. The situation changed when the first dispersion relation analysis of $\Box_{\gamma Z}$ was introduced in Ref.\cite{Gorchtein:2008px}. The dispersive representation of Eq.\eqref{eq:gammaZbox} (after removing the point-like proton contribution) reads:
\begin{eqnarray}
	\Box_{\gamma Z}^V(E)&=&\frac{2\alpha}{\pi}\int_0^\infty\frac{dQ^2}{1+Q^2/M_Z^2}\int_0^\infty d\nu\left\{\left[\frac{1}{2E}\ln\left|\frac{E+E_m}{E-E_m}\right|-\frac{1}{E_m}\right]\frac{F_1^{\gamma Z}}{m_N E}+\frac{F_2^{\gamma Z}}{2E\nu E_m}\right.\nonumber\\
	&&\left.\left[\left(1-\frac{Q^2}{4E^2}\right)\ln\left|\frac{E+E_m}{E-E_m}\right|+\frac{\nu}{E}\ln\frac{|E^2-E_m^2|}{E_m^2}\right]\frac{F_2^{\gamma Z}}{\nu Q^2}\right\}\nonumber\\
	\Box_{\gamma Z}^A(E)&=&\frac{\alpha (1-4s_w^2)}{\pi m_N E}\int_0^\infty dQ^2\frac{1}{1+Q^2/M_Z^2}\int_0^\infty\frac{d\nu}{\nu}\left[\ln\left|\frac{E+E_m}{E-E_m}\right|+\frac{\nu}{2E}\ln\frac{|E^2-E_m^2|}{E_m^2}\right]F_3^{\gamma Z}\ ,
\end{eqnarray}
where $E_m=(\nu+\sqrt{\nu^2+Q^2})/2$ (do not confuse with the electron end-point energy in Eq.\eqref{eq:delta12}), and the structure functions $F_i^{\gamma Z}$ are defined through Eq.\eqref{eq:Wmunu} with $a=Z$.
With these expressions, one indeed finds $\Box_{\gamma Z}^V(0)=0$ (for the inelastic contribution). However, when $E\neq 0$, we see that $\Box_{\gamma Z}^A$ is suppressed by the small factor $1-4s_w^2$ due to the electron weak coupling while $\Box_{\gamma Z}^V$ is not, so the relative importance of the latter increases with increasing $E$. The main discovery in Ref.\cite{Gorchtein:2008px} was that vector box possesses a significant dependence on the electron energy $E$, which contributes to a substantial uncertainty at $E\sim 1\ \text{GeV}$ where the $Q_\text{weak}$ experiment was performed~\cite{Qweak:2018tjf}. This discovery stimulated a number of follow-up studies~\cite{Sibirtsev:2010zg,Rislow:2010vi,Gorchtein:2011mz,Blunden:2011rd,Carlson:2012yi,Blunden:2012ty,Hall:2013hta,Rislow:2013vta,Hall:2013loa,Gorchtein:2015qha,Hall:2015loa,Gorchtein:2015naa,Erler:2019rmr}, and also led to the choice of a lower electron beam energy $E\approx 155\text{ MeV}$ in the upcoming P2 experiment in order to reduce the theoretical uncertainty in $Q_W^p$. 

Below we briefly describe the procedure for the dispersive analysis, starting with the vector box which depends on the parity-even structure functions $F_{1,2}$~\cite{Gorchtein:2011mz}. Recall that
\begin{equation}
	F_{1,2}\sim \sum_X(2\pi)^4\delta^4(p+q-p_X)\langle p|J_\text{em}^\mu|X\rangle \langle X|(J_Z)_V^\nu|p\rangle\ ,
\end{equation}
the two currents in the tensor are different despite that the initial and final states are the same, so it is again not straightforward to relate it to experimental cross sections. The relation, however, can be established case-by-case for each intermediate state $X$ using isospin symmetry. To do so, we first decompose the EM current and the vector neutral weak current into isosinglet ($I=0$) and isotriplet ($I=1$) pieces, neglecting the strange component: 
\begin{equation}
	J_\text{em}^\mu=\frac{1}{6}J_0^\mu+\frac{1}{2}J_1^\mu\ ,\ (J_Z)_V^\mu=-\frac{1}{3}s_w^2 J_0^\mu+\left(\frac{1}{2}-s_w^2\right)J_1^\mu\ ,
\end{equation}
where 
\begin{equation}
	J_0^\mu=\bar{u}\gamma^\mu u+\bar{d}\gamma^\mu d\ ,\ J_1^\mu=\bar{u}\gamma^\mu u-\bar{d}\gamma^\mu d\ .
\end{equation}
Next, assuming isospin symmetry, the proton, neutron and any intermediate state $X$ can be expressed as isospin eigenstates:
\begin{equation}
	|p\rangle =|1/2,1/2\rangle\ ,\ |n\rangle=|1/2,-1/2\rangle\ ,\ |X\rangle=|I_X,I_{3X}\rangle\ .
\end{equation}
With these, the transition matrix element of isospin currents between a nucleon state and an intermediate state $X$ can be expressed in terms of reduced matrix elements in the isospin space using the Wigner-Eckart theorem:
\begin{equation}
	\langle I_X,I_{3X}|J_I^\mu|1/2,\pm 1/2\rangle =C_{1/2,\pm 1/2;I,0}^{I_X,I_{3x}}\langle I_X||J_I^\mu||1/2\rangle\ ,
\end{equation} 
where $=C_{1/2,\pm 1/2;I,0}^{I_X,I_{3x}}$ is the Clebsch-Gordan coefficient. The two reduced matrix elements $\langle X||J_{0,1}^\mu||1/2\rangle$ can be fixed by the proton and neutron real~\cite{Armstrong:1971ns,Caldwell:1973bu,Caldwell:1978yb,ZEUS:2001wan,Vereshkov:2003cp} and virtual~\cite{JeffersonLabHallCE94-110:2004nsn,Whitlow:1991uw,E665:1996mob,H1:2000muc} photoabsorption cross section: $\sigma_{p\gamma^{(*)}\rightarrow X}$,  $\sigma_{n\gamma^{(*)}\rightarrow X}$, which allows a data-driven analysis of $\Box_{\gamma Z}^V$. 

In contrast with its vector counterpart, $\Box_{\gamma Z}^A$ only possesses a mild dependence on $E$, and its dispersive analysis is analogous to that of the single-nucleon $\gamma W$-box detailed in Sec.\ref{sec:DRneutron}~\cite{Blunden:2011rd,Erler:2019rmr}. First, in the perturbative region ($Q^2>2\text{ GeV}^2$), the integrand is well-described by pQCD-corrected sum rules. However, unlike the case of $\Box_{\gamma W}^V$ that depends only on the isosinglet EM current, here we need both the isosinglet and isotriplet EM current. The former satisfies the Bjorken sum rule, while the latter satisfies the Gross-Llewellyn-Smith (GLS) sum rule~\cite{Gross:1969jf}. The pQCD correction to the latter is also known to $\mathcal{O}(\alpha_s^4)$~\cite{Baikov:2010iw,Baikov:2010je,Baikov:2012zn}. The Born contribution is given by the proton magnetic Sachs form factor and the nucleon axial form factor:
\begin{equation}
	F_{3}^{\gamma Z,B}(x,Q^2)=-2G_M^p(Q^2)G_A(Q^2)\delta(1-x)\ .
\end{equation}
The $N\pi$ contribution is computed using chiral perturbation theory and is numerically small, while the leading resonance contribution from the $\Delta$ can be fixed using the transition form factor parameterization in ~\cite{Lalakulich:2005cs}. Finally, the high-energy (small $x$) contribution is again described by the Regge picture in Fig.\ref{fig:Regge}, except that now it is the $Z$-boson that fluctuates into an $a_1$, and the Regge trajectories are neutral and couple to the proton. The parameters in the Regge model are directly taken from that in the $\Box_{\gamma W}$ box analysis. Following the procedure above, Ref.\cite{Erler:2019rmr} deduced the result of the axial $\gamma Z$ box at two beam energies, corresponding to the P2 and $Q_\text{weak}$ experiment:
\begin{equation}
	\Box_{\gamma Z}^A(155\text{ MeV})=44.6(2.1)\times 10^{-4}\ ,\ 	\Box_{\gamma Z}^A(1.165\text{ GeV})=39.7(2.2)\times 10^{-4}\ .
\end{equation}

\subsection{Lattice QCD analysis}

Similar to $\Box_{\gamma W}$, the proton $\gamma Z$-box diagram $\Box_{\gamma Z}(E)$ can also be computed with lattice QCD as a four-point correlation function, which was first performed in Ref.\cite{Zhang:2026utt}. It turns out that, the $E$-dependence greatly complicates the problem, which we briefly describe as follows. Since lattice QCD operates on Euclidean spacetime, one needs to first perform a Wick rotation to the integrals in Eq.\eqref{eq:gammaZbox} to the imaginary energy domain, $q^0=i\nu_E$. The electron propagator in Eq.\eqref{eq:propagators} has two poles: 
\begin{equation}
	q_0=E-|\vec{l}-\vec{q}|+i\epsilon\ ,\ q_0=E+|\vec{l}-\vec{q}|-i\epsilon\ .
\end{equation}
In the $\ell\rightarrow 0$ limit, the pole positions reduce to $\pm(|\vec{q}|-i\epsilon)$, which are in the second and fourth quadrant and do not affect the Wick rotation; this is the case for the $\gamma W$-box. However, if $\ell$ is retained, then the pole $E-|\vec{l}-\vec{q}|+i\epsilon$ can move to the first quadrant if $\nu_e\equiv E-|\vec{l}-\vec{q}|>0$, and the Wick rotation will pick up this pole. As a result, the full integral splits into two pieces:
\begin{equation}
	\Box_{\gamma Z}(E)=\Box_{\gamma Z}^\text{Wick}(E)+\Box_{\gamma Z}^\text{res}(E)\ .\label{eq:gammaZsplit}
\end{equation}
The Wick-rotated term simply amounts to replacing $q_0$ by $i\nu_E$:
\begin{equation}
	\Box_{\gamma Z}^\text{Wick}(E)=i\alpha\int\frac{d^3\vec{q}}{(2\pi)^3}\int_{-\infty}^{\infty}\frac{d\nu_E}{2\pi}\sum_{i=1}^{3}V_i(i\nu_E,\vec{q},\ell)T_i^{\gamma Z}(i\nu_E,\vec{q})\ ,
\end{equation}
whereas the other term is contributed by the residue of the electron propagator in the first quadrant:
\begin{equation}
	\Box_{\gamma Z}^\text{res}(E)=i\alpha\int\frac{d^3\vec{q}}{(2\pi)^3}\Theta(\nu_e)\sum_{i=1}^3\text{Res}[V_i(q,\ell)]T_i^{\gamma Z}(\nu_e,\vec{q})\ ,
\end{equation}
where
\begin{equation}
	\text{Res}[V_i(q,\ell)]=\lim_{\nu\rightarrow\nu_e}(\nu-\nu_e)V_i(q,\ell)\ .
\end{equation}

The lattice treatment of the ``Wick'' and ``residue'' term in Eq.\eqref{eq:gammaZsplit} are very different. They all start with the Euclidean hadronic function $\mathcal{H}_{\mu\nu,E}(t_E,\vec{x})$ (and its spatial Fourier transofrm) involving Euclidean current operators, which is directly computable on lattice:
\begin{equation}
	\mathcal{H}_{\mu\nu}(t_E,\vec{x})=\langle p|T\{J_{\mu,E}^\text{em}(t_E,\vec{x})J_{\nu,E}^Z(0)\}|p\rangle\ ,\ \tilde{\mathcal{H}}_{\mu\nu,E}(t_E,\vec{q})=\int d^3\vec{x}e^{-i\vec{q}\cdot\vec{x}}\mathcal{H}_{\mu\nu,E}(t_E,\vec{x})\ .
\end{equation}
For the ``Wick'' component, what we need is just an ordinary temporal Fourier transform:
\begin{equation}
	T_{\mu\nu,E}(i\nu_E,\vec{q})=\int dt_E e^{-i\nu_E t_E}\tilde{\mathcal{H}}_{\mu\nu,E}(t_E,\vec{q})\ ,
\end{equation}
which treatment is exactly the same as in the $\gamma W$-box. On the other hand, the ``residue'' contribution involves a Laplace transform:
\begin{equation}
		T_{\mu\nu}(\nu_e,\vec{q})=\int_{-T/2}^{T/2} dt_E e^{\nu_e t_E}\tilde{\mathcal{H}}_{\mu\nu}(t_E,\vec{q})\ ,
\end{equation}
where a temporal extent $T$ is introduced. For any non-zero $E$, there exists intermediate states with energy than the external state energy $E+m_N$, which causes the Laplace transform to diverge as $T\rightarrow \infty$; such divergence has to be removed using the infinite-volume reconstruction method~\cite{Fu:2022fgh,Wang:2023omf,Fu:2024gxq,Tuo:2024bhm}. In addition, as $E$ increases, the location of the pole $\nu_e$ could lie above the $N\pi$ production threshold which leads to power-law finite-volume effects. Also, the long-distance reconstruction receives contributions from the $N\pi$ states. Overcoming all the obstacles above, Ref.\cite{Zhang:2026utt} reported the result at $E=155\text{ MeV}$, relevant to the P2 experiment, as:
\begin{equation}
	\Box_{\gamma Z}(E=155\text{ MeV})=5.76(32)_\text{stat}(26)_\text{ES}(11)_\text{FV}\times 10^{-3}\ ,
\end{equation}
where the uncertainties are statistical (stat), excited-state contamination (ES), and finite-volume effects (FV), respectively. This result is systematically smaller than the outcome of the dispersive analysis~\cite{Gorchtein:2015naa,Erler:2019rmr}, where the main difference comes from the vector box diagram ($1.84(31)\times 10^{-3}$ vs $2.94(18)\times 10^{-3}$). The origin of such discrepancy is yet to be understood.

\section{Conclusions}

Hadronic and nuclear uncertainties that enter the two-gauge-boson exchange diagram corrections to experimental measurables in weak interaction processes, such as beta decay lifetime and asymmetry in PVES, have to be properly taken into account to improve their sensitivity to BSM physics, and this requires non-perturbative treatments of strong interaction physics. Up to single-nucleon level, this can be done through data-driven analysis based on dispersion relation or lattice QCD calculations of four-point correlation functions. For nuclear systems, the above should be supplemented by nuclear \textit{ab initio} methods to either compute nuclear Green's functions or effective two-body potentials. Finally, for correlation coefficients in beta decays, the proper treatment of kinematics in radiative corrections, especially the bremsstrahlung process, is crucial for the extraction of the axial-to-vector coupling ratio $\lambda$.


\begin{ack}[Acknowledgments]%
 The work of C.-Y.S. is supported in
part by the DOE Topical Collaboration ``Nuclear Theory for New Physics'', award No. DE-SC0023663, and by University of Tennessee, Knoxville. 
\end{ack}

\bibliographystyle{elsarticle-num}
\bibliography{reference}

\end{document}